\documentstyle[12pt]{article}
\input{tcilatex}

\begin{document}

\title{Nonrelativistic Shifted - $l$ Expansion Technique for Three- and
Two-Dimensional Schr\"odinger Equation}
\author{Omar Mustafa \\
%EndAName
Department of Physics, Eastern Mediterranean University\\
G. Magusa, North Cyprus, Mersin 10 - Turkey\\
Thabit Barakat \\
Department of Civil Engineering, Near East University\\
Lefko{\c s}a, North Cyprus, Mersin 10 - Turkey\\
}
\maketitle

\begin{abstract}
{\small The shifted - $l$ expansion technique (SLET) has been developed to
solve for the eigenvalues of Schr\"odinger equation in three  (3D) and two
dimensions (2D). SLET simply consists of using $1/\bar{l}$ as a perturbation
parameter, where $\bar{l} = l - \beta$. $\beta$ is a suitable shift, $l$ is
the angular momentum quantum number for the 3D-case, $l=|m|$ for the
2D-case, and m is the magnetic quantum number. Unlike the shifted large - N
expansion theory (SLNT), SLET seems to be applicable  to a wider number of
problems of significant interest in physics.}
\end{abstract}

\renewcommand{\baselinestretch}{1.7} \newpage

\renewcommand{\thesection}{\Roman{section}}

\section{Introduction}

A large number of problems of physics require solutions of Schr\"odinger,
Klein-Gordon, and Dirac wave equations. Exact solutions of these equations
exist only for a handful of potentials, the fact that has culminated into
the development of many fascinating approximation techniques. However,
highly accurate analytical solutions of these equations for potentials of
arbitrary coupling constants are hard to find.

In this work we shall consider the Schr\"odinger equation for  spherically
(three-dimensional, 3D) and cylindrically (two-dimensional, 2D)  symmetric
potentials. Among the possible solutions of this wave equation in  such
symmetries exist, for example, the coupling constant perturbation theory, 
the variational method, the WKB approximation, the two-step procedure [1],
the  operator method [2], the $1/N$ expansion technique [3], the rational
fraction  approach [4], the direct numerical integration [5-7], the
supersymmetric  quantum-mechanics-based methods [8], the Hill determinant
method [9], the  multiple-step recursion relations procedure [10], and the
shifted large-N  (or $1/N$) expansion technique (SLNT) [11-13].

SLNT has surpassed the other approximations in its domain of  applicability,
effectively it puts no constraint on the potential or the  quantum numbers
involved. Moreover, its analytical expressions have yielded  very accurate
and fast converging results [11-16]. SLNT has also shown  that even if the
results of physical interest are in two [12] or in three  dimensions
[13,14], it is advantageous to work in N-dimensions and use $1/\bar{k}$ as a
perturbation parameter. $\bar{k}=N+2l-a$, N being the  number of spatial
dimensions of interest, $l$ the angular quantum number,  and $a$  is a
suitable shift which has the meaning of additional degree of freedom  and is
responsible for speeding up the convergence of the resulting energy  series.
However, the scope of its applicability to relativistic problems,  in
particular, to atomic, molecular, and quarkonium physics, has not been 
adequately explored. This is, most probably, because of the coupling nature 
of the Dirac equation that causes difficulties in inflating the  dimensions
of the mentioned equation [14,15]. The Schr\"odinger  equation for 3D
hydrogenic state in a uniform electric field is yet  another example of the
many problems that cannot be solved using  SLNT. Therefore,  in spite of its
success in the Schr\"odinger equation for spherically  [11] and 
cylindrically [12] symmetric potentials of arbitrary coupling constants, 
the shifted $1/N$ expansion has to be reformatted to widen its domain of 
applicability.

In this paper a new technique is introduced to solve the Schr\"odinger
equation. What we shall call the shifted-$l$ expansion technique (SLET) is a
reformation  of SLNT and can be easily extended in a straight forward manner
to  solve for the Dirac, the Klein-Gordon, and for any other wave equation
that can be reduced to the form of the Schr\"odinger  equation below,
Eq.(1). SLET simply consists of using $1/\bar{l}$ as  an expansion
parameter, where $\bar{l}=l-\beta$. $l$ is the angular  momentum quantum
number  for the spherically symmetric (3D) problems, $l=|m|$ for
cylindrically symmetric (2D) problems, and m is the magnetic quantum number. 
$\beta$ is a suitable shift which is mainly introduced to avoid the trivial 
case $l=0$. In addition, $\beta$ is chosen so that the next leading 
contribution to the energy eigenvalue series vanishes. This choice is 
physically motivated by requiring the agreement between the shifted-$l$ 
expansions and the exact analytical results for the harmonic-Oscillator and 
the Coulomb potentials both in 2D and 3D cases [11-13].  We shall not adopt
the classical limit, $l\rightarrow\infty$, to workout the  leading term  in
the energy series, as Imbo and co-workers [11] have done. We shall fix $\bar{%
l}$ and determine $r_{0}$ by minimizing the leading term of the energy 
eigenvalue series [12-15]. Like SLNT, SLET is also a pseudoperturbative 
technique in the sense that it proposes a perturbation parameter that is 
not directly related to the coupling constant in the potential of interest.

In Sec.II we discuss the method for spherically symmetric potentials of 
arbitrary coupling constants and give the resulting analytical  expressions
in  such away that allows the reader to use them without proceeding into
their  derivation. Inspired by the considerable attention that has been
given to  systems of reduced dimensions [12,17-21], we also discuss the
method for  cylindrically symmetric potentials of arbitrary coupling
constants in  Sec. III. In Sec.IV we shall show that the analytical results
of SLET for the power-law and logarithmic potentials coincide with those of
SLNT [11].  In the same section we shall also demonstrate that the
contribution of each  term in the energy series, Eq.(6), of SLET is the same
as the corresponding  one of SLNT [11]. In Sec.V  we shall consider a 2D
electron gas in the $x-y$ plane in the presence of a  hydrogenic potential
and a magnetic field in the z-direction. Therein, we compare our results
with those of Zhu et al [20]. Finally, we conclude and remark in Sec. VI.

\section{ SLET for the 3D Schr\"odinger equation}

To formulate SLET we start with the radial part of Schr\"odinger equation (
in $\hbar = 2m = 1$ units)\newline
\begin{equation}
\left[-d^{2}/dr^{2}+l(l+1)/r^{2}+V(r)\right]\Psi(r)=E\Psi(r).
\end{equation}
where $V(r)$ is an arbitrary spherically symmetric potential. If the angular
momentum quantum number $l$ is shifted through the relation $\bar{l} = l -
\beta$, Eq.(1) becomes\newline
\begin{equation}
\left\{-\frac{d^{2}}{dr^{2}}+\frac{\left[\bar{l}^{2}+(2\beta+1)\bar{l}
+\beta(1+\beta)\right]}{r^{2}}+V(r)\right\}\Psi(r)=E\Psi(r).
\end{equation}
It is convenient to shift the origin of coordinate by the definition [3,11]%
\newline
\begin{equation}
x=\bar{l}^{1/2}(r-r_{o})/r_{o},
\end{equation}
and to expand about $x = 0$ to obtain\newline
\begin{equation}
(\frac{x}{\bar{l}^{1/2}}+1)^{-2}=1-\frac{2x}{\bar{l}^{1/2}} +\frac{3x^{2}}{%
\bar{l}}-\frac{4x^{3}}{\bar{l}^{3/2}}+\cdots,
\end{equation}
\begin{equation}
V(x(r))=\frac{\bar{l}^{2}}{Q}\left[V(r_{o}) +V^{^{\prime}}(r_{o})\frac{r_{o}x%
}{\bar{l}^{1/2}} +V^{^{\prime\prime}}(r_{o})\frac{r_{o}^{2}x^{2}}{2\bar{l}}
+V^{^{\prime\prime\prime}}(r_{o})\frac{r_{o}^{3}x^{3}}{6\bar{l}^{3/2}}+\cdots%
\right].
\end{equation}
We also propose the expansion of E as [15]\newline
\begin{equation}
E=\frac{\bar{l}^{2}}{Q}\left[E_{o}+E_{1}/\bar{l}+E_{2}/\bar{l}^{2} +E_{3}/%
\bar{l}^{3}+\cdots\right],
\end{equation}
where $Q$ is a scale to be determined below. With the above expansions
Eq.(2) becomes\newline
\begin{eqnarray}
&&\left\{-\frac{d^{2}}{dx^{2}}+\left[\bar{l}+(2\beta+1) +\frac{\beta(1+\beta)%
}{\bar{l}}\right] \left(1-\frac{2x}{\bar{l}^{1/2}}+\frac{3x^{2}}{\bar{l}} -%
\frac{4x^{3}}{\bar{l}^{3/2}}+\cdots\right)\right.  \nonumber \\
&&  \nonumber \\
&&\left.+\frac{r_{o}^{2}\bar{l}}{Q} \left[V(r_{o})+V^{^{\prime}}(r_{o})\frac{%
r_{o}x}{\bar{l}^{1/2}} +V^{^{\prime\prime}}(r_{o})\frac{r_{o}^{2}x^{2}}{2%
\bar{l}} +V^{^{\prime\prime\prime}}(r_{o})\frac{r_{o}^{3}x^{3}}{6\bar{l}%
^{3/2}} +\cdots\right]\right\} \Phi_{n_{r}}(x)  \nonumber \\
&&  \nonumber \\
&&=\xi_{n_{r}}\Phi_{n_{r}}(x),
\end{eqnarray}
\newline
where\newline
\begin{equation}
\xi_{n_{r}}=\frac{r_{o}^{2}\bar{l}}{Q}\left[E_{o}+E_{1}/\bar{l} +E_{2}/\bar{l%
}^{2}+E_{3}/\bar{l}^{3}+\cdots\right].
\end{equation}
On the other hand, the Schr\"odinger equation for a one dimensional
anharmonic oscillator has been discussed in detail by Imbo et al. [11],
which in turn implies that\newline
\begin{eqnarray}
\xi_{n_{r}}&&=\bar{l}\left(1+\frac{r_{o}^{2}V(r_{o})}{Q}\right) +\left[%
2\beta+1+(n_{r}+\frac{1}{2})w\right]  \nonumber \\
&&+\frac{1}{\bar{l}}\left[\beta(\beta+1)+\alpha_{1}\right] +\frac{\alpha_{2}%
}{\bar{l}^{2}} +\cdots.
\end{eqnarray}
\newline
Comparing Eq.(8) with Eq.(9) and equating terms of same order in $\bar{l}$
one obtains \newline
\begin{equation}
E_{o}=Q/r_{o}^{2}+V(r_{o}),
\end{equation}
\begin{equation}
E_{1}=\frac{Q}{r_{o}^{2}}\left[2\beta+1+(n_{r}+1/2)w\right],
\end{equation}
\begin{equation}
E_{2}=\frac{Q}{r_{o}^{2}}\left[\beta(\beta+1)+\alpha_{1}\right],
\end{equation}
\begin{equation}
E_{3}=\frac{Q}{r_{o}^{2}}\alpha_{2},
\end{equation}
where $\alpha_{1}$ and $\alpha_{2}$ are given in the appendix. Here, $r_{o}$
is chosen to minimize $E_{o}$ [13-14,18], i. e.;\newline
\begin{equation}
dE_{o}/dr_{o}=0 ~~~~and~~~~d^{2}E_{o}/dr_{o}^{2}>0,
\end{equation}
\newline
which gives, with $\bar{l}=\sqrt{Q}$,\newline
\begin{equation}
l-\beta=\sqrt{r_{o}^{3}V^{^{\prime}}(r_{o})/2}.
\end{equation}
Where $\beta$ is fixed by the requirement that $E_{1}$ vanishes, then 
\newline
\begin{equation}
\beta=-[2+(2n_{r}+1)w]/4,
\end{equation}
with\newline
\begin{equation}
w=2\sqrt{3+r_{o}V^{^{\prime\prime}}(r_{o})/V^{^{\prime}}(r_{o})}.
\end{equation}
Eq.(15) is explicit in $r_{o}$ and is the same equation as that given by
Imbo et al. [11] to solve for $r_{o}$.

Finally, the eigenvalues are calculated by\newline
\begin{equation}
E=E_{o}+\frac{1}{r_{o}^{2}}\left[\beta(1+\beta)+\alpha_{1}\right] +\frac{%
\alpha_{2}}{\bar{l}r_{o}^{2}},
\end{equation}
where $n_{r}$ is the radial quantum number. and $\bar{l}$ is as much as $%
\bar{k}/2$ of Imbo et al. [11]

\section{SLET for the 2D Schr\"odinger equation}

The radial Schr\"odinger equation for cylindrically symmetric potentials can
be reduced (in $\hbar = 2m = 1$ units) to the form 
\begin{equation}
\left[-d^{2}/d\rho^{2}+\frac{(4l^{2}-1)}{4\rho^{2}}+V(\rho)\right]%
\Phi(\rho)= E\Phi(\rho).
\end{equation}
where $\rho^{2}=x^{2}+y^{2}$, $l=|m|$, and m is the magnetic quantum number
. If $l$ is shifted through the relation $\bar{l} = l - \beta$, Eq.(19)
becomes\newline
\begin{equation}
\left[-\frac{d^{2}}{d\rho^{2}}+\frac{(\bar{l}+\beta-1/2)(\bar{l} +\beta+1/2)%
}{\rho^{2}}+V(\rho)\right]\Phi(\rho)=E\Phi(\rho).
\end{equation}

Following the procedure described in Sec.II, Eqs. (3)-(6), one obtains 
\begin{eqnarray}
&&\left\{-\frac{d^{2}}{dx^{2}}+\left[\bar{l}+2\beta+\frac{(\beta^{2}-1/4)} {%
\bar{l}}\right] \left(1-\frac{2x}{\bar{l}^{1/2}}+\frac{3x^{2}}{\bar{l}} -%
\frac{4x^{3}}{\bar{l}^{3/2}}+\cdots\right)\right.  \nonumber \\
&&  \nonumber \\
&&\left.+\frac{\rho_{o}^{2}\bar{l}}{Q} \left[V(\rho_{o})+V^{^{\prime}}(%
\rho_{o})\frac{\rho_{o}x}{\bar{l}^{1/2}} +V^{^{\prime\prime}}(\rho_{o})\frac{%
\rho_{o}^{2}x^{2}}{2\bar{l}} +\cdots\right]\right\} \chi_{n_{\rho}}(x) 
\nonumber \\
&&  \nonumber \\
&&=\lambda_{n_{\rho}}\chi_{n_{\rho}}(x),
\end{eqnarray}
\newline
where\newline
\begin{equation}
\lambda_{n_{\rho}}=\frac{\rho_{o}^{2}\bar{l}}{Q}\left[E_{o}+E_{1}/\bar{l}
+E_{2}/\bar{l}^{2}+E_{3}/\bar{l}^{3}+\cdots\right].
\end{equation}
Also $\lambda_{n_{\rho}}$ can be written as [11]

\begin{eqnarray}
\lambda_{n_{\rho}}&&=\bar{l}\left(1+\frac{\rho_{o}^{2}V(\rho_{o})}{Q}\right)
+\left[2\beta+(n_{\rho}+\frac{1}{2})w\right]  \nonumber \\
&&+\frac{1}{\bar{l}}\left[(\beta^{2}-1/4)+\alpha_{1}\right] +\frac{\alpha_{2}%
}{\bar{l}^{2}} +\cdots.
\end{eqnarray}
\newline
Comparing Eq.(23) with Eq.(22) we obtain \newline
\begin{equation}
E_{o}=Q/\rho_{o}^{2}+V(\rho_{o}),
\end{equation}
\begin{equation}
E_{1}=\frac{Q}{\rho_{o}^{2}}\left[2\beta+(n_{\rho}+1/2)w\right],
\end{equation}
\begin{equation}
E_{2}=\frac{Q}{\rho_{o}^{2}}\left[(\beta^{2}-1/4)+\alpha_{1}\right],
\end{equation}
\begin{equation}
E_{3}=\frac{Q}{\rho_{o}^{2}}\alpha_{2}.
\end{equation}
Where $\rho_{0}$ is given through the condition in Eq.(14) yielding 
\begin{equation}
l-\beta=\sqrt{\rho_{o}^{3}V^{^{\prime}}(\rho_{o})/2},
\end{equation}
\begin{equation}
\beta=-\frac{1}{2}(n_{\rho}+1/2)w,
\end{equation}
\begin{equation}
w=2\sqrt{3+\rho_{o}V^{^{\prime\prime}}(\rho_{o})/V^{^{\prime}}(\rho_{o})},
\end{equation}
and $n_{\rho}$ is the radial quantum number, $n_{\rho}=0, 1, 2,\cdots$.
Finally, the eigenvalues are 
\begin{equation}
E=E_{o}+\frac{1}{\rho_{o}^{2}}\left[(\beta^{2}-1/4)+\alpha_{1}\right] +\frac{%
\alpha_{2}}{\bar{l}\rho_{o}^{2}},
\end{equation}
where $E_{0}$ is given in Eq.(24).

\section{Applications to the 3D case}

For the sake of comparison with the results of the shifted large - N
expansion technique SLNT [11], the power - law, $V(r) = A r^{\nu}$, and the
logarithmic, $V(r) = A \ln(r/b)$, potentials are considered. The above
potentials have, however, been used in heavy quarkonium spectroscopy
[11,15-17]. Our results are given in such a way that the comparison with
SLNT is made clear.

\subsection{ Power - Law Potential $V(r) = A r^{\protect\nu}$}

For the power - law potential, $V(r) = A r^{\nu}$, the general formalism
developed above leads to \newline
\begin{equation}
r_{o}=\left[\frac{2\bar{l}^{2}}{A\nu}\right]^{\frac{1}{\nu+2}},
\end{equation}
where\newline
\begin{equation}
w=2\sqrt{\nu+2}
\end{equation}
\begin{equation}
\bar{l}=[(2n_{r}+1)\sqrt{\nu+2}+(2l+1)]/2,
\end{equation}
\begin{equation}
E_{o}=\left[2A\nu\right]^{\frac{2}{\nu+2}}\frac{(4\bar{l})(\nu+2)}{8\nu} %
\left[2\bar{l}\right]^{\frac{\nu-2}{\nu+2}},
\end{equation}
\begin{equation}
\frac{E_{2}}{\bar{l}^{2}}=-\left[2A\nu\right]^{\frac{2}{\nu+2}} \frac{%
2(\nu+1)(\nu+2)}{12^{2}(2\bar{l})}\left[2\bar{l}\right]^{\frac{\nu-2} {\nu+2}%
},
\end{equation}
and\newline
\begin{eqnarray}
\frac{E_{3}}{\bar{l}^{3}}&=&\left[2\bar{l}\right]^{\frac{\nu-2}{\nu+2}} %
\left[2A\nu\right]^{\frac{2}{\nu+2}} \left[\frac{2(\nu+1)(\nu-2)}{12^{3}(2%
\bar{l})^{2}\sqrt{\nu+2}}\right]  \nonumber \\
&&  \nonumber \\
&&\times[(\nu+1)(\nu-2)+(7\nu^{2}-31\nu-62)n_{r}  \nonumber \\
&&  \nonumber \\
&&+(5\nu^{2}-29\nu-58)(3n_{r}^{2}+2n_{r}^{3})].
\end{eqnarray}
\newline

It is worth mentioning that the exact energy eigenvalues for the harmonic -
oscillator, $B^{2}r^{2}/4$, and the Coulomb, $-2/r$, potentials are obtained
from the leading term,$E_{o}$, in Eq.(6) as $E = B(2n_{r}+l+3/2)$ and $E =
-1/(n_{r}+l+1)^{2}$, respectively, where higher - order terms vanished
identically.

\subsection{ Logarithmic Potential $V(r) = A \ln(r/b)$}

For the logarithmic potential, $V(r) = A \ln(r/b)$, one may simply obtain
the following results\newline
\begin{equation}
r_{o}=\bar{l}\sqrt{2/A},
\end{equation}
where \newline
\begin{equation}
w=2\sqrt{2}
\end{equation}
\begin{equation}
\bar{l}=[(2l+1)+(2n_{r}+1)\sqrt{2}]/2,
\end{equation}
\begin{equation}
E_{o}=A\left[\ln\left(\frac{\bar{l}}{b\sqrt{A/2}}\right)+\frac{1}{2}\right],
\end{equation}
\begin{equation}
\frac{E_{2}}{\bar{l}^{2}}=\frac{A}{72\bar{l}^{2}}(6n_{r}^{2}+6n_{r}+1),
\end{equation}
and\newline
\begin{equation}
\frac{E_{3}}{\bar{l}^{3}}=\frac{A}{864\bar{l}^{3}\sqrt{2}}
(58n_{r}^{3}+87n_{r}^{2}+31n_{r}+1).
\end{equation}

For both of the potentials above, it should be pointed out that each term in
Eq.(6) has the same contribution as that of the corresponding one in SLNT
[11]. Mathematically speaking, $(E_{o})_{SLET} = (E_{o})_{SLNT}$, $(E_{2}/%
\bar{l}^{2})_{SLET} = (E_{2}/\bar{k}^{2})_{SLNT}$, and $(E_{3}/\bar{l}%
^{3})_{SLET} = (E_{3}/\bar{k}^{3})_{SLNT}$.

As found previously [11], Eqs.(22)-(24) along with Eq.(6) yield remarkably
good results even for large values of $n_{r}$. It can easily be checked that
the rate of convergence of the three terms in Eq.(6) is approximately the
same as $n_{r} \rightarrow \infty$ as it is for $n_{r} \rightarrow 0$. So,
if the results for small $n_{r}$ are accurate, one expects roughly similar
accuracy for all $n_{r}$. For more details on the accuracy of the above
predictions, Eqs.(34)-(37) for the power - law and Eqs.(40)-(43) for the
logarithmic potentials, the reader may refer to Imbo et al [11].

\section{Application to the 2D case}

In this section we shall consider the potential that describes a 2D
hydrogenic donor in the presence of a magnetic field of arbitrary strength
applied perpendicular to the 2D plane. If we adopt the symmetric gauge {\bf {%
A}=(B/2)$(-y,x,0)$, }we can express this potential as (in cylindrical polar
coordinates) [12,20,21] 
\begin{equation}
V(\rho )=-2/\rho +m\gamma +\gamma ^{2}\rho ^{2}/4.
\end{equation}
Where m is the magnetic quantum number and the magnetic field strength is
given through the parameter $\gamma $; $\gamma \sim B$. The units of energy
and length are the effective Rydberg and the effective Bohr radius,
respectively. For more details on the potential in Eq.(44) and the
parameters involved in it, one may refer to references [12,17-21]. 

Considering Eq.(31) along with Eq.(24) and (28)-(30), we have obtained the
well known limiting values of the energies of such system at the zero and
high-magnetic-field limits [23] as 
\begin{equation}
E_{donor}=-(n_{\rho }+|m|+1)^{-2},
\end{equation}
and 
\begin{equation}
E_{Landau}=\gamma (2n_{\rho }+|m|+m+1),
\end{equation}
respectively. Herein, we shall report that $E_{donor}$ and $E_{Landau}$ are
obtained by the leading term $E_{0}$ of Eq.(31), where higher-order terms
have vanished identically, i.e.; $E_{2}=E_{3}=0$. 

For the donor state in an arbitrary magnetic field, we have numerically
solved Eq.(28) through Eqs.(29)-(30) and Eq.(24) to find these states by
Eq.(31). 

In Fig.1 (to be supplied by authors), our results (curve of long dashes
marked with solid circles) for the 1s state show excellent agreement with
the other results in the weak-field regime. In the strong-field regime our
results fall in between the results of the perturbation treatment [24]
(curve of small dashes) and the direct numerical integration [23] (solid
curve) on which Zhu et al [20] have located their predictions (solid
circles). Our results are, however, unique in that they tend to approach the
strong- and weak-field perturbation theory results. The perturbation theory
coupling constants were appropriately defined in these regimes [24].
Likewise, we believe, should be the tendency of the results of any
approximation technique. 

Fig.2 (to be supplied by authors) shows the ground 1s and the $2p_{-}$
states being weakly affected by the magnetic field since the Coulomb
interaction dominates over the magnetic interaction for low-lying states.
The higher excited states, on the other hand, are more weakly bound and in
this case the parabolic quantum well, which is formed by the magnetic field,
determines the energy spectrum. Our results (solid curves marked with solid
circles) for the 1s, $2p_{-}$, and $2p_{+}$ agree with those of the series
expansion calculations [20] (solid curves). Whereas, the results for higher
excited states converge more rapidly to Landau levels (curve of small
dashes) than those of Zhu et al [20], especially in the strong-field regime
wherein the parabolic quantum well is the dominating interaction that
determines the energy spectrum. 

\section{\bf Conclusions and Remarks}

In this paper, the shifted - $l$ expansion technique (SLET) has been
developed to solve 3D and 2D Schr\"{o}dinger equations. The development is a
reformation of the shifted large - N expansion technique (SLNT) that widens
its domain of applicability. 

For the spherically symmetric (3D) power - law and the logarithmic
potentials, both SLET and SLNT have yielded the same analytical results. The
observations $(E_{o})_{SLET}=(E_{o})_{SLNT}$, $(E_{2}/\bar{l}%
^{2})_{SLET}=(E_{2}/\bar{k}^{2})_{SLNT}$, and $(E_{3}/\bar{l}%
^{3})_{SLET}=(E_{3}/\bar{k}^{3})_{SLNT}$ are significant for the conclusion
that SLET is indeed a reformation of SLNT. The accuracies and the speeds of
convergence of both techniques are the same, at least for the power - law
and the logarithmic potentials. 

For cylindrically symmetric (2D) potentials, we have considered a 2D Coulomb
field in the presence of a magnetic field of arbitrary strength, Eq.(44).
SLET results have appeared to be fast converging in the sense that the
dominating contribution to the energy series, Eq.(6), is that of the leading
term $E_{0}$.The results were compared with those of the perturbation [24],
the direct numerical integration [23], and the series expansion [20] methods
according to whichever was available. Figs.1 and 2 show that SLET results
seem more appropriate than the others. Fig.1 shows that they approach the
perturbation theory results at weak- and strong- field limits, and Fig.2
shows that they approach the Landau levels as one goes to higher-excited
states where the magnetic interaction is more effective in determining the
energy spectrum. 

In the near absence of highly accurate analytical approximation methods to
solve the Schr\"{o}dinger equation, even for simple cases such as a 3D
hydrogenic state in a uniform electric field [25] (to be investigated in the
near future), SLET provides highly accurate analytical expressions.
Moreover, the difficulties associated with the application of SLNT to Dirac
equation [14,15] should vanish for SLET. To appear elsewhere, we have
applied SLET to Dirac, and Klein-Gordon wave equations. Finally, we should
like to point out that the energy states of excitons in a
harmonic-quantum-dot [26], and of shallow donors and heavy - hole excitons
in quantum well in the presence of magnetic field can be correctly obtained
by SLET and compared with others [19,23]. 

\begin{center}
{\bf Appendix }
\end{center}

Although some of the following definitions can be found in some other
references we would like to repeat them so as to make this article self -
contained. The definitions of $\alpha _{1}$ and $\alpha _{2}$ and the
parameters involved are as follows\newline
\begin{eqnarray}
\alpha _{1} &=&[(1+2n_{r})e_{2}+3(1+2n_{r}+2n_{r}^{2})e_{4}]  \nonumber \\
&&  \nonumber \\
&-&w^{-1}[e_{1}^{2}+6(1+2n_{r})e_{1}e_{3}+(11+30n_{r}+30n_{r}^{2})e_{3}^{2}],
\end{eqnarray}
\begin{eqnarray}
\alpha _{2} &=&(1+2n_{r})d_{2}+3(1+2n_{r}+2n_{r}^{2})d_{4}  \nonumber \\
&&  \nonumber \\
&+&5(3+8n_{r}+6n_{r}^{2}+4n_{r}^{3})d_{6}  \nonumber \\
&&  \nonumber \\
&-&w^{-1}[(1+2n_{r})e_{2}^{2}+12(1+2n_{r}+2n_{r}^{2})e_{2}e_{4}+2e_{1}d_{1} 
\nonumber \\
&&  \nonumber \\
&+&2(21+59n_{r}+51n_{r}^{2}+34n_{r}^{3})e_{4}^{2}+6(1+2n_{r})e_{1}d_{3} 
\nonumber \\
&&  \nonumber \\
&+&30(1+2n_{r}+2n_{r}^{2})e_{1}d_{5}+6(1+2n_{r})e_{3}d_{1}  \nonumber \\
&&  \nonumber \\
&+&2(11+30n_{r}+30n_{r}^{2})e_{3}d_{3}+10(13+40n_{r}+42n_{r}^{2}+28n_{r}^{3})e_{3}d_{5}]
\nonumber \\
&&  \nonumber \\
&+&w^{-2}[4e_{1}^{2}e_{2}+36(1+2n_{r})e_{1}e_{2}e_{3}+8(11+30n_{r}+30n_{r}^{2})e_{2}e_{3}^{2}
\nonumber \\
&&  \nonumber \\
&+&24(1+n_{r})e_{1}^{2}e_{4}+8(31+78n_{r}+78n_{r}^{2})e_{1}e_{3}e_{4} 
\nonumber \\
&&  \nonumber \\
&+&12(57+189n_{r}+225n_{r}^{2}+150n_{r}^{3})e_{3}^{2}e_{4}]  \nonumber \\
&&  \nonumber \\
&-&w^{-3}[8e_{1}^{3}e_{3}+108(1+2n_{r})e_{1}^{2}e_{3}^{2}+48(11+30n_{r}+30n_{r}^{2})e_{1}e_{3}^{3}
\nonumber \\
&&  \nonumber \\
&+&30(31+109n_{r}+141n_{r}^{2}+94n_{r}^{3})e_{3}^{4}],
\end{eqnarray}
with\newline
\begin{equation}
e_{j}=\frac{\varepsilon _{j}}{w^{j/2}}\mbox{~~ and ~~}d_{i}=\frac{\delta _{i}%
}{w^{i/2}}\,,
\end{equation}
where $j=1,2,3,4$, $i=1,2,3,4,5,6$. 

$\varepsilon _{j}\,^{\prime }s$ and $\delta _{i}\,^{\prime }s$ for 3D case
are given as\newline
\begin{equation}
\varepsilon _{1}=-2(2\beta +1){~~,~~}\varepsilon _{2}=3(2\beta +1), 
\nonumber
\end{equation}
\begin{equation}
\varepsilon _{3}=-4+\frac{r_{o}^{5}V^{^{\prime \prime \prime }}(r_{o})}{6Q}{%
~~,~~}\varepsilon _{4}=5+\frac{r_{o}^{6}V^{^{\prime \prime \prime \prime
}}(r_{o})}{24Q},
\end{equation}
\begin{equation}
\delta _{1}=-2\beta (1+\beta ){~~,~~}\delta _{2}=3\beta (1+\beta ), 
\nonumber
\end{equation}
\begin{equation}
\delta _{3}=-4(2\beta +1){~~,~~}\delta _{4}=5(2\beta +1),  \nonumber
\end{equation}
\begin{equation}
\delta _{5}=-6+\frac{r_{o}^{7}V^{^{\prime \prime \prime \prime \prime
}}(r_{o})}{120Q}{~~,~~}\delta _{6}=7+\frac{r_{o}^{8}V^{^{\prime \prime
\prime \prime \prime \prime }}(r_{o})}{720Q}.
\end{equation}

For the 2D case $\varepsilon _{j}\,^{\prime }s$ and $\delta _{i}\,^{\prime }s
$ are given by\newline
\begin{equation}
\varepsilon _{1}=-4\beta {~~,~~}\varepsilon _{2}=6\beta ,  \nonumber
\end{equation}
\begin{equation}
\varepsilon _{3}=-4+\frac{\rho _{o}^{5}V^{^{\prime \prime \prime }}(\rho
_{o})}{6Q}{~~,~~}\varepsilon _{4}=5+\frac{\rho _{o}^{6}V^{^{\prime \prime
\prime \prime }}(\rho _{o})}{24Q},
\end{equation}
\begin{equation}
\delta _{1}=-2(\beta ^{2}-1/4){~~,~~}\delta _{2}=3(\beta ^{2}-1/4), 
\nonumber
\end{equation}
\begin{equation}
\delta _{3}=-8\beta {~~,~~}\delta _{4}=10\beta ,  \nonumber
\end{equation}
\begin{equation}
\delta _{5}=-6+\frac{\rho _{o}^{7}V^{^{\prime \prime \prime \prime \prime
}}(\rho _{o})}{120Q}{~~,~~}\delta _{6}=7+\frac{\rho _{o}^{8}V^{^{\prime
\prime \prime \prime \prime \prime }}(\rho _{o})}{720Q}.
\end{equation}
{\bf %\newpage
}

\end{document}